\documentclass[aps,showpacs,preprintnumbers,amsmath,amssymb]{revtex4}

\usepackage{graphicx}
\usepackage{dcolumn}
\usepackage{bm}
\begin{document}

\title{Non-linear plane perturbation in
a \\ non-ohmic/ohmic fluid interface in a vertical electric field}

\author{Francisco Vega Reyes}
\altaffiliation[Also at ]{Department of Physics, Georgetown University%
\\37th and O Streets NW, Washington DC, 20057, USA}
\affiliation{Departamento de Electr\'onica y
Electromagnetismo, Universidad de Sevilla. Avda Reina Mercedes
s/n. Sevilla 41018. Espa\~na}%


\date{March 14, 2005}

\begin{abstract}
The stability of a non-ohmic/ohmic fluid interface in the presence
of a constant gravitational field and stressed by a vertical
stationary electric field with unipolar injection is studied,
focusing on the destabilising action of the electric pressure when
charge relaxation effects can be ignored. We use a hydraulic
model, whose static equilibrium condition is written and analysed
as a function of the ohmic fluid conductivity when subjected to a
non-linear perturbation. The combined action of the polarization
and free interfacial charges on the pressure instability mechanism
is also analysed. The results show some important peculiarities of
the fluid interface behaviour in the presence of a stationary
space charge distribution generated by unipolar injection in the
non-ohmic fluid.
\end{abstract}

\maketitle

\section{Introduction}
\label{intro}


If a stationary electric field ${\mathbf E}$, parallel to a
constant gravitational field ${\mathbf g}$, is applied on a system
composed by two immiscible fluids with different mass densities,
the interface between them should rest, in the state of
equilibrium, completely plane and perpendicular to the fields and
eventually subjected to the destabilization when the corresponding
electric field is strong enough. In general, in this paper, the
term non-ohmic/ohmic, for example, refers to an interface where
the lower fluid layer is ohmic.

Taylor and McEwan \cite{Taylor} studied the static equilibrium of this system in the
case of a non-conducting/conducting ohmic interface and determined
the instability criterion in a linear theory. In this case, the
stress acting on the interface is acting only in its normal
direction: we say that the instability is due to a pressure
mechanism. Melcher \& Smith \cite{MelcherPhys} studied the stability of an
ohmic/ohmic interface stressed by a vertical electric field in a
more general linear theory considering all possible conductivity
values of both fluids (and also other physical properties
involved, such as viscosity, etc.), including charge relaxation
effects under these conditions. In this work, shear stresses are
involved and hence overstability \cite[]{Chandra} and surface
charge convection may occur. We say that in this case the
instability is due to the convective mechanism. (Please note that
in this case convection is due to surface charge not to volume
charge, like in the electrohydrodynamic instability due to
unipolar injection in an insulating liquid layer
\cite[]{AttenEHD,Lacroix}).

Some recent work on this problem has investigated a
non-ohmic/ohmic fluid interface when the non-ohmic layer is
subjected to unipolar injection \cite*[]{AttKoulDC,EHD}. These
works are motivated because in certain experimental systems an
electrode may act not only as a voltage source (a surface charge
source, in the end) but also as a space charge source
\cite[]{AttenEHD}. Melcher \& Schwartz \cite{MelcherTan} noticed that an electrode may
cause, if in contact with a very low conducting fluid,
dielectrical breakage and generate a stationary electric field
with a space charge distribution in the non-ohmic part of the
system. This makes the coupling of the electric field with
mechanical fluid equations very different from that occurring in
the classical studies of ohmic/ohmic fluid interfaces. A clear
example is the experiment by Koulova-Nenova, Malraison \& Atten \cite*{AttKoulBer}, where a moderate
injection in a liquid mixture of ciclohexane with TiAP salt
\cite*[]{Denat} was produced. They observed that unipolar
injection from the electrode may produce not only convection in
the bulk of the fluid but also an interfacial instability similar
to that occurring in the absence of space charge \cite[]{Taylor}
but with a peculiarity: the voltage thresholds for instability are
systematically reduced by 1/3. The complete linear theory for a
non-ohmic/ohmic interface is presented in the previous work by
Vega \& P\'erez \cite{EHD}, where a transition region in the critical behaviour of
the interface has been found. This region marks the
conducting-to-insulating transition in the behaviour of the
non-ohmic/ohmic interface. In fact, the existence of this
transition region implies that the dynamics of the same
non-ohmic/ohmic interface subjected to unipolar injection may
behave like in an ohmic/ohmic interface in which the lower layer
is the most conducting, but also like in an ohmic/ohmic interface
in which the lower layer is the least conducting (insulating
behaviour). However, by definition, in the non-ohmic/ohmic
interface the lower layer is always the most conducting. The
apparent contradiction comes from the fact that, under unipolar
injection, the electric conduction in the non-ohmic layer may be
actually more "effective" than the ohmic conduction in the lower
layer, depending on the value of the applied electric potential.
This causes the mechanics of the fluid interface to be much more
complex when there is injection.

The purpose of this paper is to demonstrate that this complexity
appears already in the electric pressure instability mechanism;
i.e., the static interfacial equilibrium between electric and
gravitational forces. In order to make clear an intuitive
visualization of this equilibrium, a hydraulic model is developed
(figure \ref{figsetup}). The system described in the model is not
that of an infinite fluid interface and thus the results are not,
in general, quantitatively applicable to the infinite interface
problem. However, as it will be shown, the results of the
hydraulic model can account for the same transition region
described above \cite[]{EHD}. Thus, the hydraulic model yields a
qualitatively analogous description of the corresponding problem
in the infinite interface. But the results are not only restricted
to comparison of results in a previous work \cite[]{EHD}. They
also provide the following additional inputs: a) identification of
the mechanism that causes the transition region to appear
(electric pressure due to polarization charges) and precise
determination of the transition region, and b) observation of this
transition as a function of the perturbation amplitude (not only
as a function of conductivity as previously detected
\cite[]{EHD}). As we will see this implies that, once the
instability begins, an interface with a very conducting lower
layer may be stabilized in states with non-zero perturbation
amplitudes. This behaviour differs from that observed in the
purely ohmic case \cite[]{Taylor,AttKoulBer}, where the
interfacial perturbation grows continually because the electrical
pressure mechanism is always actively pulling up or pushing down
the interface.


The interest of this problem is due to the original behaviour of
this putative fluid interface and its possible industrial
applications. For example, the formation of stable metallic liquid
points when the interface changes from conducting (the electric
pressure has opposite sign to the applied field) to non-conducting
behaviour (the electric pressure keeps the same sign that the
applied field) in a perturbed state could be used to make ion
sources.

The possibility of producing ion sources by manipulating a fluid
interface with electric fields motivated the work by
N\'eron de Surgy \cite{ThNeron}, who extended the original work by Taylor and McEwan \cite{Taylor}
introducing a non-linear perturbation in a
non-conducting/conducting ohmic fluid interface, but always
without injection. The results proved theoretically that a
metallic liquid can never develop stable points, independently of
the geometry of the system. However,
 in some rare cases he experimentally observed stable metallic points,
 which N\'eron indicated could be due to
impurities in the liquid. We suggest in this paper that
 this is related to an injection from the metallic liquid points to the air.



In \S\ref{syshydr} we describe the hydraulic model and find its
non-linear stability condition. In \S\ref{comparacion} we find the
difference between the stability condition in the hydraulic model
and the one found for the infinite interface in a previous work
\cite*[]{EHD}. The effect of the combined action of polarization
and free surface charges will be explained in \S\ref{diagramas}.
In \S\ref{condiciones} we introduce the reduced critical parameter
$U_{NL}$ and the possible general behaviours of the non-linear
critical curves are described. Finally, in \S\ref{oh/oh} and
\S\ref{no-oh/oh} we present the results of the hydraulic model in
the cases of an ohmic/ohmic interface and the non-ohmic/ohmic
interface, respectively. Although the purely ohmic interface has
been studied extensively \cite*[]{Taylor,MelcherEHD,ThNeron}, the
results of \S\ref{oh/oh} are interesting to make evident the new
perspective gained with the hydraulic model.

\section{Hydraulic model}
\label{hidraulico}

\subsection{The system and the equilibrium equation}
\label{syshydr}

\begin{figure}
\includegraphics[height=8.5cm]{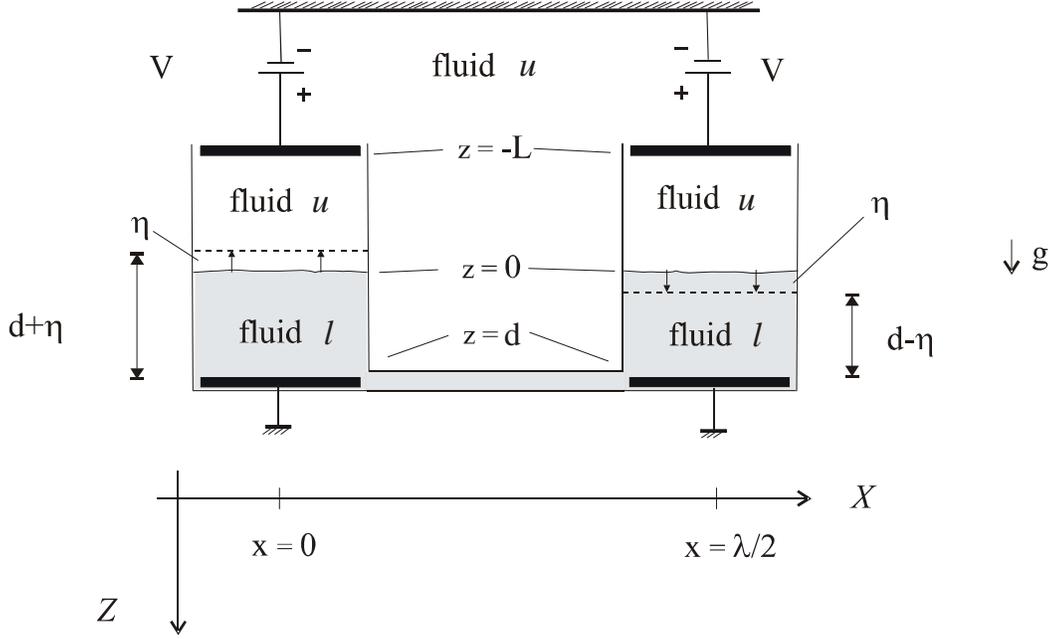}
\caption{The system studied. A plane perturbation of amplitude
$\eta$ is introduced in such a way that the interface level is
raised in the cylinder centered in $x=0$ and lowered in the other
cylinder (which is centered at $x=\lambda/2$).} \label{figsetup}
\end{figure}

The system (figure \ref{figsetup}) consists of two identical rigid
cylinders with parallel axes, connected to each other through a
thin horizontal cylindrical pipe at their base. The system is in a
constant gravitational field with a constant acceleration
$\mathbf{g}=g\mathbf{u}_z$ (this field acts in a direction called
"vertical"; thus, its perpendicular plane defines the horizontal
directions). The system is immersed in a fluid \textit{u}, which
supplies a constant pressure on another fluid (we call it fluid
\textit{l}), which is denser than fluid \textit{u}. Both fluids
are immiscible and incompressible, so in the equilibrium state the
fluid \textit{l} layer is below the other one. The radius $r$ of
the horizontal thin pipe is large enough to allow a negligible
Poiseuille effect for any typical fluid velocity
 (i.e., $\Delta p /(8\mu l v/r^2)\ll 1$, where $\mu$ is
the dynamic viscosity, $l$ is the pipe length, $v$ the fluid
velocity and $\Delta p$ is a typical pressure difference in the
vertical direction) and thus the pressure from a vertical cylinder
is completely communicated to the other one. There is a pair of
horizontal rigid electrodes in each cylinder, one at the top of
the cylinder  (at $z=-L$), and the other at its base (at $z=d$).
The length of the system ($L+d$) is much lower than the horizontal
dimensions of the cylinders so the boundary effects are
negligible. The upper electrodes are connected to the same DC
voltage source, that supplies a voltage $V$, while the lower
electrodes are grounded. Additionally, if fluid \textit{u} is
non-ohmic, a space charge source at the upper electrodes injects
unipolar volume charge $q_0$ at $z=-L$.

We write the Navier-Stokes equation:
 \begin{equation} \rho \frac{d{\mathbf v}}{dt}=-\nabla p+\mu\nabla^2{\mathbf
v}+\nabla\cdot{\mathcal T}^e+\rho{\mathbf g} \label{NS}
\end{equation} where $\rho$ is the mass density, $\mathbf{v}$ the
fluid velocity, $p$ the total pressure and ${\mathcal T}^e$
denotes the electric stress tensor, whose elements are ${\mathcal
T}^e_{ij}=\varepsilon E_iE_j-\frac{1}{2}\delta_{ij}\varepsilon
E^2$ ($\varepsilon$ is the dielectric constant of the fluid,
subscripts $i$ and $j$ indicate the components in cartesian
coordinates ($i$,$j$=$x$, $y$, $z$) of the stress tensor and the
electric field of modulus $E$, and $\delta_{ij}$ are the elements
of the identity tensor).

We define the jump of a magnitude $A$ as the difference between
its values just below and just above the interface and denote it
as $<A>=A_l(F(\mathbf{r}))-A_u(F(\mathbf{r}))$, where
$F(\mathbf{r})$ is the interface position. The normal stress
balance condition in the interface is written:


\begin{equation}
\mathbf{n}\cdot\left<\mathcal{T}^v+\mathcal{T}^e\right>\mathbf{n}-\left<p\right>=\gamma\nabla\cdot\mathbf{n}
\end{equation} where $\mathbf{n}$ is the normal direction to the
interface, $\gamma$ is the coefficient of surface tension and
$\mathcal{T}^v $ is the viscous stress tensor for incompressible
fluids: $\mathcal{T}^v_{ij}=\mu\left(\partial v_i/\partial
j+\partial v_j/\partial i\right)$ (again, $i$,$j$=$x$, $y$, $z$).
From now on, we will ignore coordinate $y$, due to the system
symmetry.



 If the system is in equilibrium, the fluid interface is
horizontal and at the same level in both cylinders (figure
\ref{figsetup}). In this situation all stresses at the interface
are in the vertical direction and we have $\mathbf{v}=\mathbf{0}$
and $\nabla\cdot\mathbf{n}=\nabla\cdot\mathbf{u}_z=0$. Then the
normal stress balance reads:

\begin{equation}
\left<p\right>=\left<\frac{1}{2}\varepsilon E^2\right>
\label{Nsalto}
\end{equation}




We introduce now a plane perturbation and the interface level is
raised a height $\eta$ in one cylinder (the one at $x=0$) and
lowered the same height $\eta$ in the other (Fig. \ref{figsetup}).
The perturbation is kept by some pressure source until the
electric field becomes stationary. In this point, we still have
$\mathbf{v}=\mathbf{0}$ and
$\nabla\cdot\mathbf{n}=\nabla\cdot\mathbf{u}_z=0$. Now the
pressure source stops and the system is left under the action of
gravitational and electric pressures. It is implicit in the way in
which the perturbation is introduced that at this point charge
relaxation may be ignored and that condition (\ref{Nsalto}) is
still fulfilled. We express now the pressure as a function of the
scalar field $\Pi$, which is defined by the relation:
$\Pi=p-\rho\mathbf{g}\cdot\mathbf{r}$ and that is called "modified
pressure" \cite{Batchelor}. The equation (\ref{NS}) in an
equilibrium state may be written then:

\begin{equation}
\nabla \Pi=\nabla\cdot{\mathcal T}^e \label{Pivol}
\end{equation} which reflects the fact that the total pressure
that $\Pi$ the surface a body immersed in the fluid feels is $p$
modified by the gravitational force \cite{Batchelor}. When the
gradient of this modified pressure $\Pi$ is in balance with
electric stresses in the volume of the fluid, like in
(\ref{Pivol}), there is no net volume force in our system. Respect
to the net pressure jump at the interface, it can be rewritten as
a function of the modified pressure $\Pi$:

\begin{equation}
\left<p\right>_--\left<\rho\right>g\eta=\left<\Pi\right>_-, \quad
\quad
\left<p\right>_+-\left<\rho\right>g(-\eta)=\left<\Pi\right>_+
\end{equation} where subscript "$-$" stands for the value of the
magnitude at the interface in the cylinder at $x=\lambda/2$
(downward perturbation) and subscript "$+$" stands for the value
of the magnitude at the interface in $x=0$ (upward perturbation).
If we use these expressions into (\ref{Nsalto}) and we take the
difference between the pressure jumps in both cylinders we obtain:

\begin{equation}
\left<\Pi\right>_--\left<\Pi\right>_++2\left<\rho\right>g\eta=
\left<\frac{1}{2}\varepsilon E^2 \right>_-
-\left<\frac{1}{2}\varepsilon E^2\right>_+ \label{Neq}
\end{equation}

When the modified pressure jump at the interface in both cylinders
is the same, the surface forces at the interface in both cylinders
are equilibrated. Thus, our stability condition is:

\begin{equation}
\left<\Pi\right>_--\left<\Pi\right>_+=0 \label{equilibrio}
\end{equation}

This is the only mechanical equation we are actually using in this
work, from now on. When the electric pressure term overcomes the
gravitational term then $\left<\Pi\right>_--\left<\Pi\right>_+>0$
and the perturbation tends to amplify. If
$\left<\Pi\right>_--\left<\Pi\right>_+<0$ the perturbation tends
to damp. The stability condition leads (\ref{equilibrio}) to:

\begin{equation}
\left<\frac{1}{2}\varepsilon
E^2\right>_--\left<\frac{1}{2}\varepsilon
E^2\right>_+=2\left<\rho\right>g\eta \label{SaltoPre}
\end{equation}

In the absence of electric forces, the stability condition
(\ref{SaltoPre}) gives the solution $\eta=0$ (i.e., if no electric
field is applied, logically the only possible equilibrium state is
the same interface position in both cylinders).

It can be demonstrated that for infinitesimal $\eta$, the
hydraulic model stability condition (\ref{SaltoPre}) reproduces
the linear instability criterion for an infinite plane fluid
interface under a vertical electric field in the limit of long
wavelength if the horizontal variation of the pressure is zero, as
we will see in the following section.


\subsection{The hydraulic model and the problem of the linear stability
of an infinite plane fluid interface} \label{comparacion}

In the analogous and more studied problem of an infinite plane
fluid interface stressed by a vertical stationary electric field,
several mechanisms may simultaneously induce an
electrohydrodynamic instability, unlike in the hydraulic model
here developed, where the electric pressure is the only
destabilizing mechanism. In order to make a comparison between the
results that will be presented in this work and the results in the
problem of the infinite interface, two questions could be posed:
1) are there situations in which, like in the hydraulic model, the
electric pressure is the only active instability mechanism in an
infinite plane interface?; and 2) if so, can the hydrostatic model
account for its linear instability threshold values?

Concerning the first question \cite{EHD} demonstrated that in the
infinite plane interface the electric pressure
 is the only destabilizing mechanism involved
when an initial perturbation with infinite wavelength
($\lambda=\infty$) was produced. And this infinite wavelength
instability occurs if capillary forces are strong enough
\cite*[]{Taylor} and additionally, in the specific case of a
non-ohmic/ohmic interface, if the non-ohmic fluid has a very high
ionic mobility \cite*[]{EHD}. Assuming that we deal with an
infinite plane fluid interface whose properties fulfill these
conditions (i.e., that the long wavelength instability occurs) and
concerning question 2), the answer is yes, although not always. As
we will see, this is because in the hydrostatic model the
variation of the pressure in the horizontal is not taken into
account. This can be shown if the linear perturbation of
Navier-Stokes equation term for horizontal components $x$, $y$ are
considered. Given the symmetry of the system, it is enough to
analyse the $x$ component:


 \begin{equation} \frac{\partial \delta p}{\partial x}
 +q\frac{\partial\delta\phi}{\partial x}=
 \left(\mu\nabla^2-\rho\frac{\partial}{\partial t}\right)\delta v_x
 \label{tanpre} \end{equation} where $q$ is the free space charge in the equilibrium
 state, and $\delta\phi$
 and $\delta p$ are
  the electric potential and the total pressure linear
  perturbations in a very slightly deformed interface,
  while $\delta v_x$ is the $x$ component of the velocity linear perturbation.



Integrating this equation in $x$ and taking into account that
$\delta v_x=0$ in the limit of long wavelength \cite*[]{EHD}, when
the jump at the interface is taken the following relation is
obtained:

\begin{equation}
\left<\delta\Pi\right>=-\left<q\delta\phi\right>
\end{equation}

Consequently, the total jump of the modified pressure at the
interface is now:

\begin{equation}
\left<\Pi+\delta\Pi\right>=\left<\frac{1}{2}\varepsilon
E^2\right>-\left<\rho\right> g z_s-\left<q\delta\phi\right>
\end{equation} where $z_s$ is the interface level. Thus

The stability condition to be used for an infinite interface is
the following:

\begin{equation}
\left<\Pi+\delta\Pi\right>_--\left<\Pi+\delta\Pi\right>_+=0 \label{Geneq}
\end{equation}

In the case of an ohmic/ohmic interface the additional term is
always zero because no volume charges are present ($q=0$), which means
that the hydraulic model yields the exact linear criterion for an
infinite plane interface in the limit of long wavelength. However,
 in the non-ohmic/ohmic interface this term is not zero except
in the cases of a perfect conducting ohmic fluid,
$\sigma_l=\infty$ (being $\sigma_l$ its electric conductivity),
and a non-conducting ohmic fluid, $\sigma_l=0$,
 where the interface is an equipotential, and thus $\delta\phi=0$ in the interface.
 It is convenient to comment
 at this point that (\ref{Geneq}) is equivalent to the linear stability
  condition for the infinite interface found in an author's previous work \cite*[]{EHD}, as
  we will numerically check out in section 4. The advantage of
  starting out from the hydraulic model is evident if we notice that the deduction
  of the stability condition has
  become now much simpler \cite*[]{EHD}.

  As we see, then, the additional term $\left<\delta\Pi\right>$
 is in general needed to obtain the exact criterion for an infinite and initially
  plane non-ohmic/ohmic interface.
Nevertheless, the simpler stability condition
  (\ref{equilibrio}) obtained for the system of two cylinders not only describes
essentially, in a non-ohmic/ohmic interface under unipolar
injection, the instability regions as a function of the ohmic
conductivity but also the linear criterion threshold value, as we
will see in \S\ref{no-oh/oh}.



We recall that the comparison to the hydraulic model is restricted
to an infinite wavelength perturbation in the infinite interface.
For shorter characteristic wavelengths capillary forces are
involved but the present analysis is useful because it provides an
intuitive description on the mechanical process that occurs at the
interface due to the action of electric pressure against
gravitation. And this action is present as the main destabilizing
mechanism in any problem of a fluid interface stressed by a
vertical electric field.


\subsection{Dimensionless magnitudes. An intuitive framework}
\label{diagramas}

We take $d$, $\left<\rho\right> gd$, and $\sqrt{\left<\rho\right>
gd^3/\varepsilon_u}$ respectively as reference units for distance,
pressure, and electric potential, being $\varepsilon_u$ the
permittivity of fluid \textit{u}. From now on we will only use
non-dimensional magnitudes, and we will denote them with the same
symbols that we used for the dimensional ones, except for the
perturbation amplitude, that we will call $\xi=\eta/d$.

The static equilibrium of the interface in the hydraulic model is
given by the opposition of a gravitational term and an electric
term. In a perturbed state the gravitational term always acts
towards the part of the system with a lower thickness of the
heavier fluid layer. On the contrary, the electric pressure may
act towards any of the two cylinders, depending on both the
magnitude and sign of the electric pressure jump in each cylinder.
Thus, it is convenient to write the dimensionless electric
pressure jump in the interface using a parameter that allows to
easily determine case by case the electric pressure sign:

\begin{equation}
 \left<\frac{1}{2}\varepsilon
E^2\right>=\frac{1}{2}\varepsilon_lE_l(z_s)^2-\frac{1}{2}E_u(z_s)^2
=\frac{1}{2}E_l(z_s)^2\left(\varepsilon_l-\Sigma^2\right)
  \label{Gsalpre}
\end{equation} where we use subscripts $u$ and $l$ to denote
the magnitudes in fluids $u$ and $l$ respectively, $z_s$ is the
interface position and $\Sigma\equiv E_u(z_s)/E_l(z_s)>0$, that we
call the "apparent conductivity" of the interface. It reflects
which of the two fluids is more conducting in the interface:
$\Sigma>1$ if $E_u(z_s)>E_l(z_s)$ and we say that the interface is
conducting and conversely, not conducting if $\Sigma<1$. The
expression (\ref{Gsalpre}) is valid for all fluids independently
of their regime of electric conduction.

The total surface charge $Q_t$ and the free surface charge $Q$ at
the interface are, respectively:

\begin{equation}
Q_t=\varepsilon_0(1-\Sigma)E_l(z_s) \quad\quad
Q=(\varepsilon_l-\Sigma)E_l(z_s) \label{intcarga}
\end{equation} where $\varepsilon_0$
is the reduced vacuum permittivity. The stability condition
(\ref{SaltoPre}) in reduced magnitudes gets:
\begin{equation}
\left<\frac{1}{2}\varepsilon
E^2\right>_--\left<\frac{1}{2}\varepsilon E^2\right>_+=2\xi
\label{adeq}
\end{equation}



There are two special cases for which an eventual instability due
to electric pressure is not possible, as the left hand term in
(\ref{adeq}) is null. In effect, in the case
$\Sigma=\sqrt{\varepsilon_l}$ it is evident, from (\ref{Gsalpre}),
that each one of the electrical pressure terms in the left hand of
(\ref{adeq}) is zero so there is no non-trivial solution ($\xi\neq
0$) to this equation. And in case the equality $\left<\varepsilon
E^2\right>_-=\left<\varepsilon E^2\right>_+$ is fulfilled, the
left term of the stability condition (\ref{adeq}) is again null
and a non-trivial solution does not exist.

\begin{figure}
\includegraphics[height=8.5cm]{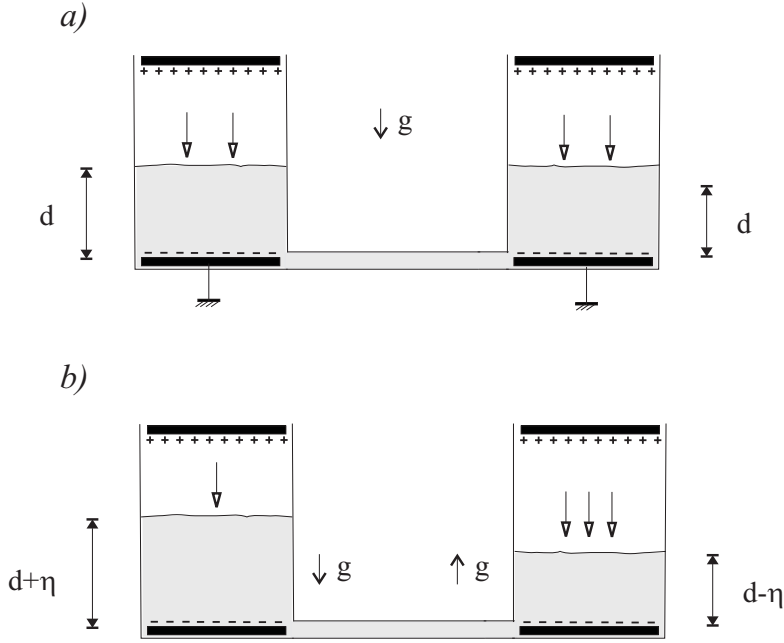}
\caption{A system with $\Sigma<\Sigma_0$ when the pressure
instability mechanism is possible ($\Sigma<\sqrt{\varepsilon_l}$).
Long white arrows indicate the action of electrical pressure jump.
In (\textit{b}), when a perturbation is introduced, we see a
gravitational force appear. This gravitational pressure flow,
directed towards the cylinder with lower interface height in all
cases, is $2\left<\rho\right>g\eta$ (or $2\xi$ in reduced
magnitudes). In this case, the net electric pressure flow is
directed towards the left cylinder (three arrows against one).}
\label{mechIn}
\end{figure}

The electric pressure jump, or equivalently (\ref{Gsalpre}),
$E_l(z_s)$, decreases with the lower layer thickness if the
apparent conductivity is high enough; i.e., the lower layer is
"less conducting". This occurs if $\Sigma<\Sigma_0$, where the
value of $\Sigma_0$ depends on the electrical regime of conduction
of the fluids. 
Figure \ref{mechIn}(\textit{a}) represents the initial unperturbed
state when the electrical pressure jump is positive
($\Sigma<\sqrt{\varepsilon_l}$) and decreases with the lower layer
thickness ($\Sigma<\Sigma_0$).

If a plane perturbation is introduced, the magnitude $\Sigma$ gets
in general a value $\Sigma_-$ in the cylinder with a minimum
interface elevation and another value $\Sigma_+$ in the cylinder
with maximum elevation. But in order to simplify, let us restrict
to an infinitesimal perturbation, so the new values still fulfill
$\Sigma_{\pm}<\Sigma_0,\varepsilon_l$. Once the perturbation is
introduced, a gravitational pressure acts against it by
communicating an upwards pressure to the zone with minimum
interface elevation (figure \ref{mechIn}\textit{b}). Now the
electric pressure jump (\ref{Gsalpre}) is higher in the zone with
a lower interface height as the thickness of the lower fluid layer
has decreased. Conversely, in the zone of maximum elevation the
electric pressure jump decreases. Thus, a net electric pressure
flow appears towards the left cylinder in figure
\ref{mechIn}(\textit{b}). If the difference between the electrical
pressure jump in both cylinders is high enough, the restoring
action of the gravitational pressure will be counterbalanced. This
is possible if the applied potential is higher than a critical
value $V_c$, given by the stability condition (\ref{equilibrio}).

\begin{figure}
\includegraphics[height=8.5cm]{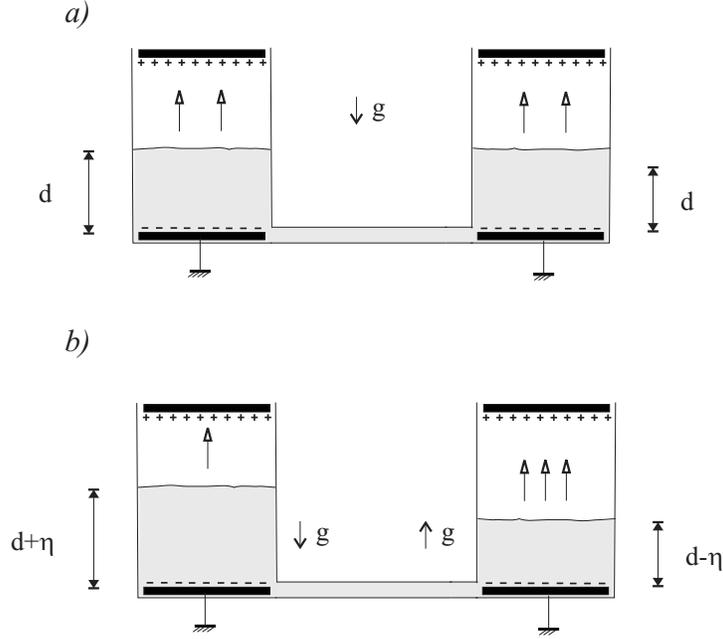}
\caption{A system with $\Sigma<\Sigma_0$ when the pressure
instability mechanism is not possible
($\Sigma>\sqrt{\varepsilon_l}$). We see now the net electric
pressure flow is directed towards the right cylinder (lower
interface height), in the same restoring direction that the
restoring gravitational term: the electric pressure is
stabilising.} \label{mechEs}
\end{figure}

On the contrary (figure \ref{mechEs}\textit{a}), if the electrical
pressure jump magnitude still decreases with the lower layer
thickness but becomes negative
($\sqrt{\varepsilon_l}<\Sigma<\Sigma_0$), the perturbation is
damped (figure \ref{mechEs}\textit{b}).
 The same analysis may be carried out
when $E_l(z_s)$ increases with the layer thickness of fluid $l$
($\Sigma>\Sigma_0$), so finally two linearly stable regions are
found: $\sqrt{\varepsilon_l}<\Sigma<\Sigma_0$
 and $\Sigma_0<\Sigma<\sqrt{\varepsilon_l}$. Thus, the apparition of
linearly stable bands is related to the change in the tendency of
the electric pressure with the lower layer thickness, which gives
the limit $\Sigma_0$, and to the change of electric pressure jump,
which gives the limit $\varepsilon_l$.

We see that for a given behaviour of $E_l(z_s)$, the stabilization
is due to a change in the sign of the electric pressure jump. In
order to find out what may cause this change of sign, let us write
the electric pressure jump as a function of polarization
($Q_p=Q_t-Q$) and free interfacial charges. The contribution of
the free surface charges to the electrical pressure jump has the
same sign of the total interfacial charge term:
\begin{equation}
\left<\frac{1}{2}\varepsilon
E^2\right>=\frac{1}{2}E_l(z_s)\left[\frac{\Sigma}
{\varepsilon_0}Q_t+Q\right]
\end{equation} while the contribution of the polarization charges has opposite
sign to the term of total charge:
\begin{equation}
\left<\frac{1}{2}\varepsilon
E^2\right>=\frac{1}{2}E_l(z_s)\left[\left(1+\frac{\Sigma}
{\varepsilon_0}\right)Q_t-Q_p\right] \label{prepol}
\end{equation} Thus, we see that if no polarization
charges are present initially, the electric pressure jump takes
the sign of the total interfacial charge $Q_t$. If we now change
$\varepsilon_l$ at constant $\Sigma$ in such a way that
polarization charges have the same sign that $Q_t$ and get high
enough, they can change the sign of the electric pressure jump.
Thus, they play an essential role in the stabilization of the
interface.


\subsection{On the critical curves in the hydraulic model}
\label{condiciones}

It is also convenient to define the dimensionless parameter
$U_{NL}=\varepsilon_uV_c^2/(\left<\rho\right> gd^3)$ where $V_c$
is the applied potential for which the stability condition
(\ref{equilibrio}) is fulfilled (i.e., $V>V_c$ yields
$\left<\Pi\right>_--\left<\Pi\right>_+>0$ and the perturbation can
be sustained). The parameter $U_{NL}$ represents the critical
electric pressure $\varepsilon_uV_c^2/d^2$, reduced with the
gravitational pressure $\left<\rho\right>gd$, while its square
root $U_{NL}^{1/2}$ represents the reduced critical applied
potential. $U _{NL}$ is in general a function of the perturbation
amplitude $\xi$.

 Let us start with an unperturbed state ($\xi=0$, $V^2<U_{NL}(\xi=0)$),
 where an arbitrary stationary perturbation with amplitude $\xi_0$ is
introduced. Then a valid solution $U_{NL}(\xi_0)$ from
(\ref{adeq}) should fulfill two conditions to make possible the
perturbation be sustained:

\textit{i)} $U_{NL}(\xi_0)>0$.

\textit{ii)} $V^2\geq U_{NL}(\xi_0)$.


The first one comes from the definition of $U_{NL}$ because
$U_{NL}(\xi_0)<0$ should correspond to an imaginary critical
potential. The second one is also necessary because $V^2\leq
U_{NL}(\xi_0)$ indicates that the perturbation is decreased to a
lower value.

Typical critical curves $U_{NL}(\xi)$ are represented in figures
\ref{figOhmico}(\textit{a}) and \ref{fig2}. In these curves stable
and unstable regions are delimited by the function $U_{NL}(\xi)$
and the behaviour of $U_{NL}(\xi)$ provides information about the
evolution of the perturbation once it is introduced. An increasing
$U_{NL}(\xi)$ in $\xi_0$ ($\partial U(\xi_0)/\partial\xi>0$) means
that the perturbation will not tend to increase for an applied
potential $V=U_{NL}^{1/2}(\xi_0)$, because for $\xi>\xi_0$,
 $V<U_{NL}^{1/2}(\xi)$ (the interface
cannot overcome the restoring action of the gravitational
pressure). And viceversa, a decreasing $U_{NL}$ in $\xi_0$ means
that the perturbation will tend to increase if
$V=U_{NL}^{1/2}(\xi_0)$. The third case occurs when $\partial
U(\xi_0)/\partial\xi=0$. In this case, for $V=U_{NL}^{1/2}(\xi_0)$
the perturbation will not increase if $\partial^2
U_{NL}(\xi_0)/\partial\xi^2>0$ but it will tend to increase if
$\partial^2 U_{NL}(\xi_0)/\partial\xi^2<0$. In definitive, if
$\partial U_{NL}(\xi_0)/\partial\xi>0$ (or if $\partial
U(\xi_0)/\partial\xi=0$, $\partial^2
U_{NL}(\xi_0)/\partial\xi^2>0$) it may be said that
"stabilization" of the perturbation occurs at
$V=U_{NL}^{1/2}(\xi_0)$ and the point $\xi=\xi_0$ may be
considered as a new point of stable interface position, that is
different to the trivial solution $\xi=0$ of the initial
equilibrium state. We call these points "perturbed stable states"
and as we will see they are only present in the non-ohmic/ohmic
interface.

\section{Ohmic/ohmic interface} \label{oh/oh}

In the case of two ohmic fluids with conductivities $\sigma_u$,
$\sigma_l$, we take $\sigma_u$ as unit for electric conductivity.
Taking into account that in the stationary state
$\nabla\cdot{\mathbf j}=0$, then $j_u=j_l=j$, $E_u=\sigma_lE_l$
and the expressions for the dimensionless stationary electric
field and $\Sigma$ in the unperturbed state are the following:
 \begin{equation} E_l=V_l \quad \quad E_u= V_u/L=\sigma_lV_l/L \quad \quad
\Sigma=\sigma_l \label{charge}\end{equation}
 being $V_l$ and $V_u$ the electric
potential drop through the lower and upper fluid layers inside the
cylinders. As the perturbation is plane, the fields in the
perturbed state are:
\begin{equation}
E_{l\pm}=\frac{V_{l\pm}}{1\pm\xi} \quad \quad
E_{u\pm}=\frac{V_{u\pm}}{L\mp\xi} \label{campoOh}
\end{equation} where the upper signs stand for the
magnitudes evaluated at $x=0$ and the lower signs stand for the
magnitudes at $x=\lambda/2$.

Using
 $V=V_{u+}+V_{l+}=V_{u-}+V_{l-}$, we obtain:
\begin{equation}
 \left<\frac{1}{2}\varepsilon
 E^2\right>_{\pm}=\frac{1}{2}\left(\varepsilon_l-\sigma_l^2\right)
 \frac{V_{l\pm}^2}{(1\pm\xi)^2}
  \label{salpre}
\end{equation}
\begin{equation}
V_{l\pm}=\frac{1\pm\xi}{\sigma_l\left(L\mp\xi\right)+\left(1\pm\xi\right)}V
\end{equation}

In an ohmic/ohmic interface $\Sigma$ is a constant (\ref{charge})
and therefore we can make the interface at constant $\Sigma$
coincide with the real interface (constant conductivities of the
fluids). It is also easy to find that $\Sigma_0=1$: $E_l(z_s)$
always decreases with the $l$ layer thickness for $\Sigma<1$ and
always increases for $\Sigma>1$ (\ref{campoOh}). Thus, given the
conductivities and permittivities of the fluids, the interface
should be stable respect to the pressure instability mechanism for
any value of the electric field, if
$\sqrt{\varepsilon_l}<\Sigma<1$ or
$1<\Sigma<\sqrt{\varepsilon_l}$, following the analysis in
\S\ref{diagramas}. This can be confirmed by finding the function
$U_{NL}$.

After some short calculations and using the stability condition in
reduced magnitudes (\ref{adeq}) we get $U_{NL}$ as a function of
the perturbation amplitude and the other parameters of the system:
\begin{equation}
U_{NL}(\xi,\sigma_l,\varepsilon_l,L)=f(\sigma_l,\varepsilon_l,L){\mathcal
K}(\xi,\sigma_l,L) \label{UNL}
\end{equation} where $f(\sigma_l,\varepsilon_l,L)$ and
${\mathcal K}(\xi,\sigma_l,L)$ (which gives all the dependence on
the perturbation amplitude $\xi$) are the following functions:
\begin{eqnarray}
\nonumber  f(\sigma_l,\varepsilon_l,L)
=\frac{1}{(\varepsilon_l-\sigma_l^2)(1-\sigma_l) (1+L\sigma_l)}  &
&
\\
 {\mathcal
K}(\xi,\sigma_l,L)=\left((1+L\sigma_l)^2-(1-\sigma_l)^2\xi^2\right)^2
& & \label{fun}
\end{eqnarray}

\begin{figure}
\includegraphics[height=6.0cm]{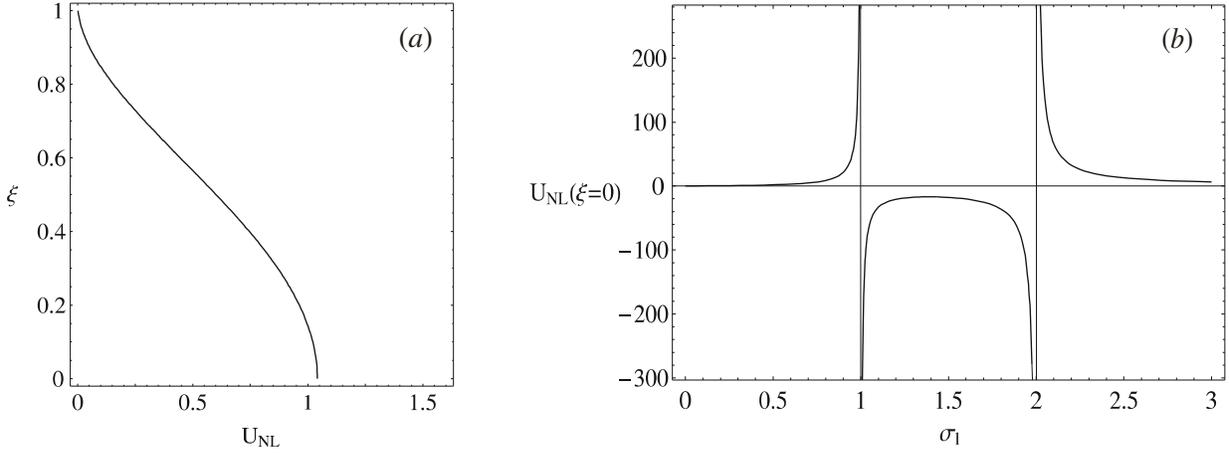}
\caption{(\textit{a}) The parameter $U_{NL}$ (if positive) is
always decreasing with the perturbation amplitude; $\sigma_l=100$,
$\varepsilon_l=4$, $L=1$. (\textit{b}) The linear criterion
($U_{NL}(\xi=0)$) plotted against $\sigma_l$ presents a stable
region (the negative values of $U_{NL}$), as a consequence of the
action of polarization charges; $\varepsilon_l=4$, $L=1$.}
\label{figOhmico}
\end{figure}

It is to be noticed that for $\Sigma=\sigma_l=1$ and
$\Sigma=\sigma_l=\sqrt{\varepsilon_l}$ the parameter $U_{NL}$
takes an infinite value for any $\xi$ (i.e., the instability is
not possible), which is in accordance with the analysis in
\S\ref{diagramas} ($\sigma_l=\sqrt{\varepsilon_l}$ is equivalent
to $\Sigma=\sqrt{\varepsilon_l}$ and $\sigma_l=1$ gives
$\left<\varepsilon E^2\right>_-=\left<\varepsilon E^2\right>_+$).

In an ohmic/ohmic interface, condition \textit{i)} in
\S\ref{condiciones} is fulfilled if and only if
$f(\sigma_l,\varepsilon_l,L)>0$. This is so because the function
${\mathcal K}(\xi,\sigma_l,L)$ is always positive (\ref{fun}). And
the sign of $f(\sigma_l,\varepsilon_l,L)$ is positive (\ref{fun})
unless the terms ($1-\sigma_l^2/\varepsilon_l$) and ($1-\sigma_l$)
have different signs. This occurs when
$\sqrt{\varepsilon_l}<\sigma_l<1$ or
$1<\sigma_l<\sqrt{\varepsilon_l}$; i.e., no perturbation can be
sustained if the electrical pressure jump and the total surface
charge $Q_t$ at the interface have opposite signs (see
(\ref{intcarga}), (\ref{charge}), (\ref{salpre})). Thus, the
result advanced by the analysis in \S\ref{diagramas} is confirmed
by the explicit calculation of $U_{NL}$. Analysing further the
conditions for stabilization we notice that they are fulfilled
when the interfacial free charge
$Q=(\varepsilon_l-\Sigma)E_l(z_s)$ takes opposite sign to the
total interfacial charge $Q_t$, and consequently the polarization
charge $Q_p=Q_t-Q$ takes the sign of $Q_t$, which agrees again
with the analysis performed in \S\ref{diagramas}. Although the
problem of the stability of an ohmic/ohmic interface has been
extensively studied, this stabilizing effect due to polarization
charges has not been formerly detected. The values of $U_{NL}$ are
plotted in figure \ref{figOhmico}(\textit{b}) vs. the reduced
conductivity $\sigma_l$. The stable region corresponds to the
interval where $U_{NL}$ takes negative values.

Besides, we saw in \S\ref{condiciones} too that for
$V^2=U_{NL}(\xi_0)$ the perturbation amplitude tends to grow up to
$\xi_1>\xi_0$ from its initial value $\xi_0$ if $V^2\geq
U_{NL}(\xi)$ inside the interval $(\xi_1,\xi_0)$, which is always
the case in an ohmic/ohmic interface if conditions \textit{i)} and
\textit{ii)} are fulfilled. In effect, let us study the derivative
$\partial U_{NL}/\partial\xi$:
\begin{eqnarray}
& & \frac{\partial
  U_{NL}}{\partial\xi}=f(\sigma_l,\varepsilon_l,L)\frac{\partial{\mathcal
K}(\xi,\sigma_l,L)}{\partial\xi}
 \\ & &  \nonumber
 \frac{\partial{\mathcal
K}(\xi)}{\partial\xi}=-4\xi(\sigma_l-1)^2\left((1+L\sigma_l)^2-(\sigma_l-1)^2\xi^2\right)
\end{eqnarray} As we see, if $f(\sigma_l,\varepsilon_l,L)>0$, $\partial
U_{NL}/\partial\xi$ is always negative (figure
\ref{figOhmico}\textit{a}), provided that $\xi<L$ (a perturbation
with $\xi>L$ forces the interface to touch the upper electrode,
case that we do not analyse). The result is independent of the
initial of the values of $\xi_0$ and $\xi_1$. This means that in
an ohmic/ohmic fluid interface the pressure mechanism is always
self-fed as it becomes increasingly stronger: once the instability
is set on the perturbation amplitude grows up to its maximum
value. This is a well known characteristic of EHD instabilities in
plane interfaces in ohmic systems \cite*[]{ThNeron}.

For $\sigma_l\to\infty$, it is easy to find from (\ref{UNL}) and
(\ref{fun}) that $U_{NL}(\xi=0)=L^3$, wich agrees with the result
by \cite{Taylor}. And finally, the author has checked that for the
linear instability at finite conductivities the critical values
provided by the hydraulic model ($U_{NL}(\xi=0)$) coincide exactly
with those of the complete linear theory for an infinite plane
interface \cite*[]{MelcherPhys} in the case of an infinite
wavelength instability with negligible charge relaxation effects,
confirming the demonstration in \S\ref{comparacion}.

We see then that although the instability in an infinite
ohmic/ohmic interface has been extensively studied, the hydraulic
model is able to reproduce some former basic results
\cite*[]{MelcherPhys,ThNeron} and also to find a new feature: the
stabilizing effect of the polarization interfacial charges when
combined with the action of free interfacial charges. But the most
relevant new results of the hydraulic model are found in the
non-ohmic/ohmic interface under unipolar injection, in the next
section.





\section{Non-ohmic/ohmic interface}
\label{no-oh/oh}

\subsection{Equations} \label{equations}

Let us suppose now that the fluid "u" is in non-ohmic regime of
electric conduction and the fluid "l" is in ohmic regime. If there
is a unipolar space charge source in the upper electrode, a
unipolar space charge distribution is induced in the non-ohmic
fluid, in which the conduction (in dimensional magnitudes) is
expressed by ${\mathbf j_u}=q_u(K_u{\mathbf E}_u+{\mathbf
v})-D\nabla q_u$; where $K_u$ is the ion mobility of non-ohmic
fluid and $D_u$ its diffusion coefficient. The diffusion term may
usually be neglected \cite*[]{AttenEHD}, so in a state at rest
(${\mathbf v}={\mathbf 0}$) we have in our system that
$j_u=K_uq_uE_u$ in the non-ohmic fluid. We take now
$K_u\sqrt{\varepsilon_u\left<\rho\right> g/d}$ as unit for
electric conductivity, being $K_u$ the ion mobility in fluid
\textit{u}. In reduced magnitudes we have the following electric
equations in stationary regime (for which $\nabla\cdot{\mathbf
j}=0$):

\begin{equation}
j_{u}=q_uE_{u}=j \quad \quad \frac{dE_u}{dz}=q_u
\end{equation}

\begin{equation}
j_l=\sigma_l E_l=j \quad \quad \frac{dE_l}{dz}=0
\end{equation} with the boundary conditions in the electrodes:

\begin{equation}
\phi_u(-L)=1  \quad  \quad q_u(-L)=C \quad \quad \phi_l(1)=0
\label{iny}
\end{equation} and in the interface:

\begin{equation}
\phi_u(0)=\phi_l(0) \quad \quad j_u(0)=j_l(0)=j
\end{equation} where the condition $q_u(-L)=C$ denotes the fact
that there is a space charge source at $z=-L$. The parameter $C$
represents the reduced space charge that the upper electrode
injects on and is usually called "level of injection". The
non-ohmic conduction due to unipolar injection and the correct
boundary condition for a unipolar injection source $q_u(-L)$ have
been studied in detail by Atten \cite*[and references therein for
more details on this issue]{AttenEHD,Lacroix}.

Then the solution of the stationary electric field is:

\begin{equation}
E_u(z)=\sqrt{2j(z+b)} \quad \quad E_l(z)=\frac{j}{\sigma_l} \quad
\quad \Sigma =\sqrt{\frac{2b}{j}}\sigma_l  \label{campoUL}
\end{equation} with $b=\frac{j}{2C^2}+L$.

We see that now $\Sigma$ is not a constant but a function of the
electric field (through the current density) and it is always
possible for any real interface (i.e., any given value of
$\sigma_l$) in the initially unperturbed state
 to find ranges of the electric
field for which its corresponding $\Sigma$ lies out of the stable
intervals: there exist real $U_{NL}(\xi=0)$ which are solution of
the stability condition (\ref{equilibrio}) for all values of
$\sigma_l$. This means that the linearly stable region around the
intervals $\sqrt{\varepsilon_l}<\Sigma<\Sigma_0$ or
$\Sigma_0<\Sigma<\sqrt{\varepsilon_l}$ only exist as a consequence
of dimensionalization and therefore they should disappear for a
real interface (constant $\sigma_l$). In any case we will see in
\S\ref{transicion} that now the action of polarization charges
affects the pressure instability mechanism in other ways.

If the system is under strong injection conditions (i.e., $C\to
\infty$), the expressions for the electric field and $\Sigma$ in
the unperturbed interface are the following:

 \begin{equation} E_l(0)=V_l \quad \quad E_u(0)=
 \sqrt{2jL}=
 \sqrt{2\sigma_lV_lL} \quad \quad
 \Sigma=\sqrt{\frac{2\sigma_lL}{V_l}} \label{sno}
  \end{equation}

If a plane perturbation of amplitude $\xi$ is introduced, the
equation for the electric pressure jump in the cylinder at the
interface is:



  \begin{equation} \left<\frac{1}{2}\varepsilon
 E^2\right>_\pm=\frac{1}{2}\left[\varepsilon_l\frac{V_{l\pm}^2}{(1\pm\xi)^2}
 -2V_{l\pm}\sigma_l\left(\frac{L\mp\xi}{1\pm\xi}\right)\right]
 \label{salto+}
  \end{equation}



The electric pressure jump can be expressed as a function of the
applied potential $V$ taking into account: $j_{u+}=j_{l+}$,
$j_{u-}=j_{l-}$ and $V=V_{u-}+V_{l-}=V_{u+}+V_{l+}$ for the
electric potential. Then, in the perturbed interface we get:



 \begin{equation} (V-V_{l\pm})^2=\beta_\pm V_{l\pm} \label{caida}
\end{equation} where $\beta_\pm=\sigma_l(L\mp\xi)^3/(1\pm\xi)$.



From (\ref{caida}), we get the solution of $V_{l\pm}$:




\begin{equation} V_{l\pm}=\frac{1}{2}\left(2V+\beta_\pm -\sqrt{\beta_\pm^2+4\beta_\pm V}\right)
\label{VVc}  \end{equation}

The other root of (\ref{caida}), the one corresponding to the sign
$+$ before the square root, is not possible as it gives a solution
of $V_{l\pm}$ such that $V_{l\pm}>V$. In effect, as
$\sqrt{\beta^2+4\beta V}> \beta$:

 \begin{equation}
 V_{l\pm}=\frac{2V+\beta_\pm+\sqrt{\beta_\pm^2+4\beta_\pm
V}}{2}>\frac{2V+\beta_\pm}{2}>V  \end{equation}

The value of $\Sigma_0$ is not constant in the non-ohmic/ohmic
interface (from (\ref{salto+}) and (\ref{VVc})) and in general
$\Sigma_0\ne 1$, unlike in the ohmic/ohmic case. We introduce the
expressions for the electric pressure jumps in the stability
condition in reduced magnitudes (\ref{adeq}) and then we get:

\begin{equation} \left(\varepsilon_l
\frac{V_{l-}^2}{(1-\xi)^2}-2\sigma_lV_{l-}\frac{L+\xi}{(1-\xi)}\right)-\left(\varepsilon_l
\frac{V_{l+}^2}{(1+\xi)^2}-2\sigma_lV_{l+}\frac{L-\xi}{(1+\xi)}\right)=4\xi
 \label{critNL}
\end{equation}




As $V_{l+}$ and $V_{l-}$ are, from (\ref{VVc}), determined by $V$,
when the equality is held in (\ref{critNL}) we get the condition
of minimum applied potential $U_{NL}^{1/2}$ for which the
instability mechanism is possible. Unlike in the ohmic/ohmic case,
now (in general) there is no analytical expression for $U_{NL}$.


\subsection{Perfect conductor limit}\label{percond}

Let us suppose that the ohmic fluid is a perfect electric
conductor ($\sigma_l\to\infty$). In this limit $V_l\to 0$ and
then:

 \begin{equation} \left<\frac{1}{2}\varepsilon E^2\right>_\pm\to\frac{1}{2}E_{u\pm}^2
~\propto~\sigma_lV_{l\pm}=\frac{\sigma_l}{2}\left(2V+\sigma_la_\pm-
\sigma_la_\pm\sqrt{1+\frac{4V}{\sigma_la_\pm}}\right)
\label{desarr}
\end{equation} where $a_\pm=(8/9)(L\mp\xi)^3/(1\pm\xi)$.

We analyse the limit of $\sigma_lV_l$ when $\sigma_l\to\infty$.
Developing the square root in power series of $1/\sigma_l<<1$, we
get from  (\ref{critNL}) and (\ref{desarr}):






 \begin{equation} U_{NL}=\frac{4}{9}\mathcal{F}(\xi) \label{NLIF}  \end{equation}
 being $\mathcal{F}(\xi)$:

 \begin{equation} {\mathcal F}(\xi)=\frac{4\xi(L^2-\xi^2)^2}{(L+\xi)^2-(L-\xi)^2}=
\frac{1}{L}\left(L^2-\xi^2\right)^2
  \end{equation}

  The function $\mathcal{F}(\xi)$ gives the variation of the
non-linear criterion with the perturbation amplitude $\xi$.

Let us compare now with the case without charge injection ($C=0$).
In the absence of injection the solution of the electric field at
the interface tends to $E_{u\pm}\to V_{u\pm}/(L\mp\xi)$. Operating
in a similar way to the strong injection case we have:




 \begin{equation} U_{NL}=\mathcal{F}(\xi) \label{NLNI}
\end{equation} that is exactly the same to the case of strong
injection (\ref{NLIF}) except for the factor $4/9$ that now does
not appear.

The dependence of $U_{NL}$ (and $U_{NL}^{1/2}$) with $\xi$ is,
from (\ref{NLIF}, \ref{NLNI}), the same for $C=\infty$ and $C=0$
and is given by the function $\mathcal{F}(\xi)$ (or
$\mathcal{F}^{1/2}(\xi)$ for $U_{NL}^{1/2}$). We can see this
behaviour of $U_{NL}^{1/2}(\xi)$ in fig. \ref{fig2}(\textit{a}):
for $\sigma_l=\infty$ (and hence, also for $\sigma_l=0$),
$U^{1/2}_{NL}$ is always a decreasing function of $\xi$. Then,
once the instability starts it tends abruptly to states
 with a minimum $U_{NL}$, which are the ones having a maximum value of the
perturbation amplitude. An analogous behaviour has been experimentally
 observed in a conducting liquid with \cite[]{AttKoulBer} and without
injection \cite[]{Taylor,ThNeron}, who observed that the
instability develops violently towards the upper electrode,
producing an electric breakage.


In the limit of zero perturbation amplitude the linear criterion
for the instability in a perfect conducting fluid interface can be
reobtained. In effect, if ${\xi\to 0}$ we obtain that ${\mathcal
F}(\xi)\to L^3$, and then $U_{NL}^{1/2}=L^{3/2}$ for the case
without injection, which agrees with previous works
\cite*[]{Taylor,ThNeron} and $U_{NL}^{1/2}=(2/3)L^{3/2}$ for the
case with infinite injection, which agrees again with the result
in former works \cite*[]{AttKoulDC,EHD}. Note also that the case
$C=0$ yields the same criterion that in the ohmic/ohmic case for
$\sigma_l\to\infty$, in \S\ref{oh/oh}.


\subsection{Non-conductor limit} \label{noncond}

Analogously, the non-conductor limit can be taken. Developing
(\ref{critNL}) in power series of $\sigma_l\ll 1$, and in the
limit of  $\sigma_l\to 0$ we get:






 \begin{equation} U_{NL}=\left(\frac{1}{\varepsilon_l}\right)
 \frac{4\xi(1-\xi^2)^2}{(1+\xi)^2-(1-\xi)^2}=
\frac{1}{\varepsilon_l}\left(1-\xi^2\right)^2=\frac{1}{\varepsilon_l}{\mathcal
H}(\xi) \label{NLpc}  \end{equation} where now before the function
of $\xi$ appears a factor $1/\varepsilon_l$ instead of the 4/9 in
the perfect conductor. The value $\sigma_l=0$ has not real
physical meaning but it is interesting the study of this limit as
a reference for very low conductivities. In definitive, the
behaviour of $U_{NL}^{1/2}(\xi)$ for $\sigma_l\ll 1$ is similar to
that in the limit $\sigma_l\to\infty$, represented in figure
\ref{fig2}(\textit{a}). This can be seen analytically in the
function ${\mathcal H}(\xi)$, which has the same behaviour that
the function ${\mathcal F}(\xi)$: $U_{NL}(\xi)$ is always
decreasing and is zero for the maximum perturbation amplitude. A
similar behaviour is detected in experiments in very low
conducting liquids under unipolar injection: the rose-window
instability has a high deformation amplitude (of the order of the
liquid layer thickness) near the instability threshold
\cite*[]{Tesis}. The peculiarity respect to the high conducting
case is that now the electric pressure is directed downwards.

The zero perturbation amplitude limit, $\xi\to 0$, yields
${\mathcal H}(\xi)=1$ and $U_{NL}^{1/2}=1/\sqrt{\varepsilon_l}$,
which agrees with the result in the linear theory for the infinite
interface \cite*[]{EHD}.


\subsection{Non-linear transition from conducting regime to low
conducting regime} \label{transicion}

In \S\ref{percond} and \S\ref{noncond}, we have demonstrated that
in the limits of a non-conductor and perfect conductor ohmic layer
$U_{NL}(\xi)$ is minimum for the maximum perturbation amplitude.
This means that the instability, once is set on, evolves up to the
maximum perturbation amplitude. The difference between both limits
comes from the fact that while in the perfect conductor the
instability is due to an upward electric pressure, in the
non-conductor limit the instability is driven by a downward
electric pressure. It seems reasonable to think that between the
perfect conductor and the non-conductor behaviour there should be
intermediate behaviours, i.e., perturbations that do not grow up
to the maximum value. In effect, we saw in \S\ref{equations} that
the real non-ohmic/ohmic interface is always initially unstable:
it is possible to find a finite value of the applied electric
field that is able to sustain any finite perturbation, which is a
consequence of the non-constancy of $\Sigma$ (\ref{sno}) in the
non-ohmic/ohmic case. But this does not prevent, after the
instability is set on, the interface from entering a stabilizing
behaviour. There would be two possibilities: one is that the
electric pressure jump changes sign, and the other is that this
pressure jump changes its behaviour with the lower layer
thickness, while the interface is evolving. Either of the two
possibilities could make the interface enter the stable intervals
$\Sigma_0>\Sigma>\sqrt{\varepsilon_l}$ and
$\sqrt{\varepsilon_l}>\Sigma>\Sigma_0$. We recall this is possible
in the non-ohmic/ohmic interface only because the apparent
conductivity (\ref{sno}) is not constant, and thus the interface
is taking new $\Sigma$ values as the perturbation evolves. Thus,
although linear stable states do not exist now for any value of
$\sigma_l$, it should be possible to think in these cases of an
instability that evolves without reaching the maximum perturbation
amplitude: we say that perturbed states are expected to appear.

\begin{figure}
\includegraphics[height=14.5cm]{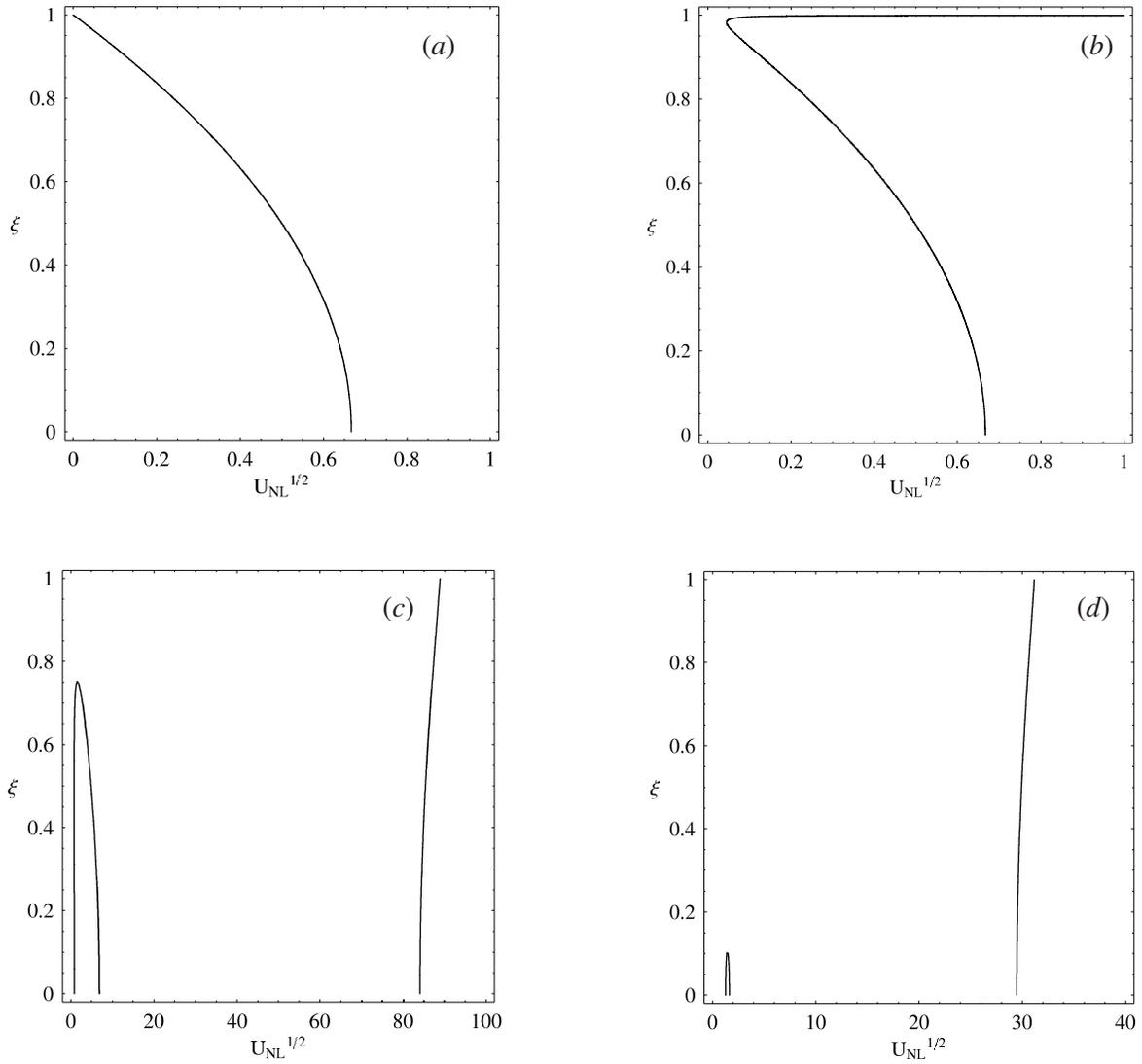}
\caption{$U_{NL}^{1/2}$ as a function of the non-linear
perturbation amplitude $\xi$ in a non-ohmic/ohmic interface under
strong injection (dimensionless $L=1$, $\varepsilon_l=4$ in all
figures). (\textit{a}) For a non-ohmic/perfect conducting
interface. (\textit{b}) For a high non-dimensional conductivity
value ($\sigma_l=10^4$). (\textit{c}) For a lower non-dimensional
conductivity ($\sigma_l=20$) three solutions in $\xi=0$ appear.
(\textit{d}) The two first solutions in $\xi=0$ tend to disappear
if the conductivity is still lowered ($\sigma_l=7$).} \label{fig2}
\end{figure}

We saw in \S\ref{condiciones} that the perturbed stable states
occur in an interval of $\xi$ when $\partial U_{NL}/\partial\xi>0$
in it. And, in effect, critical points of this type are calculated
for the first time in this work, in figure \ref{fig2}(\textit{b}),
where $U_{NL}^{1/2}$ is plotted for a non-dimensional conductivity
$\sigma_l=10^4$. This reduced conductivity corresponds in an
air/liquid interface to a physical conductivity of the order of
$10^{-2}~\Omega^{-1}$m$^{-1}$ when the dimension of the system is
of about 1 cm. We see in figure \ref{fig2}(\textit{b})that the
function $U_{NL}(\xi)$ (or, equivalently $U_{NL}(\xi)^{1/2}$)
obtained from (\ref{critNL}) has an increasing behaviour near the
maximum amplitude. This means that a fluid much more conducting
than water, for example, shows such stabilization, which could be
possible even for an initial value $\xi_0=0$ and
$V=U_{NL}^{1/2}(0$) as once the interface reaches the increasing
$U_{NL}^{1/2}(\xi)$ region it begins to be decelerated as soon as
it takes a value $\xi_1$ for which $V<U_{NL}^{1/2}(\xi_1)$.

We saw in the linear theory for an infinite plane interface that
there may exist multiple instability threshold values for a given
conductivity \cite*[]{EHD}. This also occurs in the hydraulic
model because if we further decrease the conductivity, three
solutions at $\xi=0$ begin to appear (figure
\ref{fig2}\textit{c}). Two of them enclose an unstable region
which should correspond with the high conducting regime-like
instability (upward electric pressure jump) while the third one
corresponds to the low conducting mechanism (downward electric
pressure jump). The two solutions enclosing an unstable region get
nearer as the conductivity is decreased (figure
\ref{fig2}\textit{d}): the conducting mechanism tends to
disappear. Finally, for lower conductivities, a unique solution at
$\xi=0$ corresponds to the non-conductor limit of (\ref{NLpc}).
The value of $\Sigma$ (\ref{sno}) can be used to determine the
corresponding pressure mechanism.

\begin{figure}
\includegraphics[height=6.5cm]{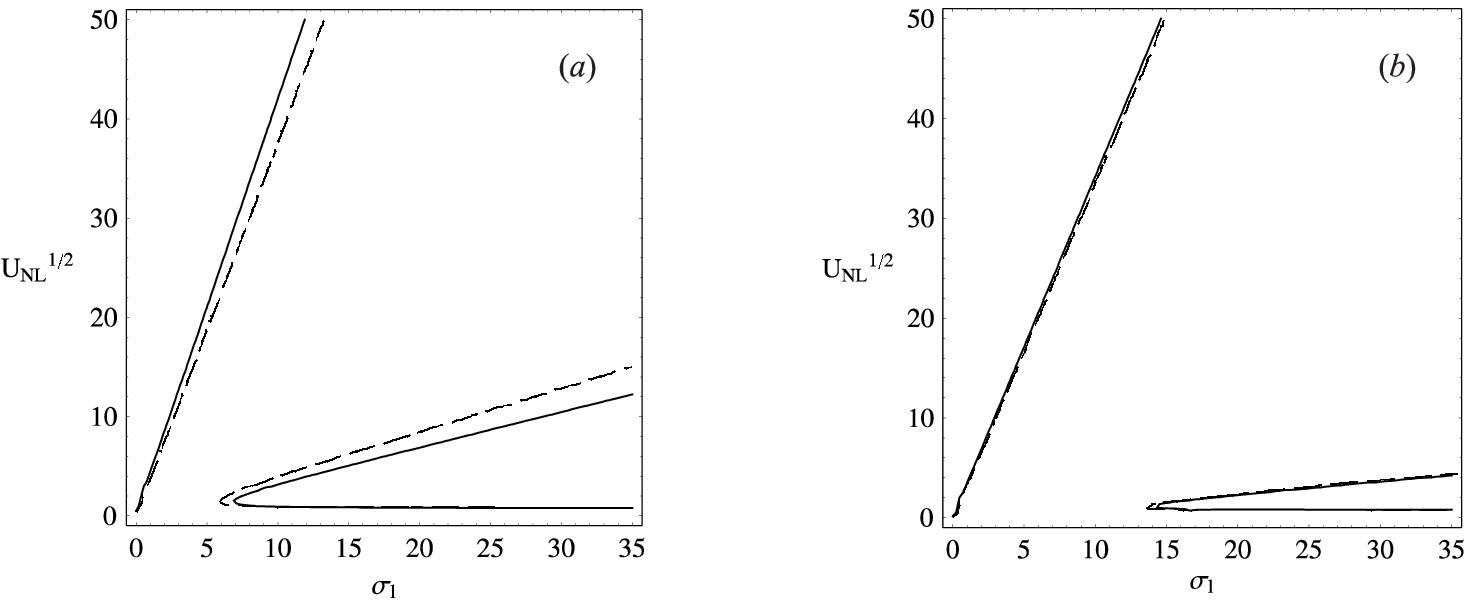}
\caption{Reduced critical potential $U_{NL}$ vs. reduced
conductivity $\sigma_l$ for an infinitesimal perturbation $\xi\to
0$ with $L=1$. Continuous curve stands for the hydraulic model
stability condition (\ref{equilibrio}) while discontinuous curve
stands for both the stability condition in a previous work and
(\ref{Geneq}). (\textit{a}) Reduced dielectric constant
$\varepsilon_l=4$. (\textit{b}) Reduced dielectric constant
$\varepsilon_l=40$. The effect of increasing the dielectric
constant (i.e., increasing the polarization interfacial charges in
the high conducting region) in the non-ohmic/ohmic interface is to
move the conducting critical curve (the one at the right) to
higher values of $\sigma_l$.} \label{Tran}
\end{figure}

Figure \ref{Tran} represents $U_{NL}^{1/2}($$\xi$$=0)$ as a
function of the conductivity $\sigma_l$. The curves from the
stability condition for the hydraulic model (\ref{equilibrio})
show the same qualitative behaviour that the linear critical
values from the stability condition for the infinite interface
(\ref{Geneq}). In addition, the linear critical values from
(\ref{Geneq}) coincide (figure \ref{Tran}) with those for an
infinite interface in the dimensional representation (scaled to
our dimensionless magnitudes) of a previous work \cite*[]{EHD},
which confirms that the stability condition (\ref{Geneq}) applies
for an infinite interface. The quantitative similarity of the
results from (\ref{equilibrio}) and (\ref{Geneq}) in figure
\ref{Tran} suggests also that additional term in the stability
condition for the infinite interface is less important. Thus, the
limits of the stable bands detected in the dimensionless
representation of the author's previous work \cite*[]{EHD} must be
close the limit values for the hydraulic model:
$\sqrt{\varepsilon_l}$ and $\Sigma_0$. Thus, the stable bands are
related to the change of electric pressure jump sign and the
change of tendency of the electric pressure jump with layers
relative thickness, instead of being related to the change of sign
in the interfacial charge, as suggested previously \cite*[]{EHD}
(notice, too, that the change of sign of interfacial charge does
not provide two limits but just one value).

\section{Conclusions}

We study a hydraulic model, for which we find a non-linear
stability condition, with a twofold objective:

First, to study specifically the unstabilizing mechanism of
electric pressure jump against the gravitational force, we use a
dimensionless representation in which we introduce the fundamental
parameter "apparent conductivity" ($\Sigma$). This allows a
systematic study of the different types of electric pressure
unstabilising mechanisms. The simple and intuitive hydraulic model
and the use of the apparent conductivity, allow us to find two
stability bands and their origins. This is in the action of
polarization charges, that may inhibit the instability pressure
mechanism in an interface subjected to a vertical field when
combined to free interfacial charges. In our formulation, the
stability bands are: $\Sigma_0>\Sigma>\sqrt{\varepsilon_l}$ and
$\sqrt{\varepsilon_l}>\Sigma>\Sigma_0$, where $\varepsilon_l$ and
$\Sigma_0$ are the reduced dielectric constant and the value of
the apparent conductivity for which the electric pressure passes
from decreasing to increasing with the reduced lower layer
thickness. As a consequence of this, the pressure instability
mechanism is not possible for any ohmic/ohmic interface with
reduced conductivity $\sigma_l$ in the intervals
$1>\sigma_l>\sqrt{\varepsilon_l}$ and
$\sqrt{\varepsilon_l}>\sigma_l>1$; i.e., the total surface charge
and the electric pressure jump at the interface have opposite
signs. In the case of a non-ohmic/ohmic interface, the pressure
instability mechanism is always possible, for any value of the
ohmic conductivity, when we pass to the representation in
$\sigma_l$.

In the hydraulic model there is also a transition region
\cite[]{EHD} in the pressure instability mechanism in a
non-ohmic/ohmic interface for a linear perturbation, but
\textit{also} for a non-linear perturbation. The behaviour of the
parameter $U_{NL}$ (or $U_{NL}^{1/2}$) as a function of the
perturbation amplitude reveals the existence, for intermediate
conductivities, of perturbed stable states (which are different to
the trivial solution of the initial equilibrium state $\xi=0$).
These new stable states are only found in the non-ohmic/ohmic
interface, suggesting the possibility of interfacial dynamics very
different to those detected in systems without injection
\cite*[]{Taylor,MelcherEHD,MelcherPhys}. This could be related to
the observations by \cite{ThNeron} of stabilised metallic points.
In the limits of very low or high conductivity the behaviour of
the function $U_{NL}^{1/2}(\xi)$ is as expected (analogous to the
observed in the infinite interface): the instability evolves up to
the maximum value of perturbation amplitude.

Although a detailed study of the interfacial dynamics is needed in
order to do determine precisely the situations in which the
stabilisation in perturbed states from any initial state is
possible, it seems evident that the introduction of the injection
enriches the behaviour of the pressure equilibria in a fluid
interface, and also opens a way to stabilization and control of
the interface deformation of high conducting fluid interfaces by
applying stationary electric fields.

The second main objective of this work is to approach, in a
mathematically simple way, the stability condition found in a
previous work for an infinite non-ohmic/ohmic interface under
unipolar injection, in the limit of long wavelength \cite*[]{EHD}.
In this issue, an equivalent stability condition to the referred
one is found in a very simple way, in \S\ref{comparacion}. The
difference between the linear stability conditions in the
hydraulic model (\ref{equilibrio}) and in the infinite interface
(\ref{Geneq}) was also found. This allows us to determine if there
are cases in which these conditions coincide. We found that this
coincidence occurs in the case of an ohmic/ohmic interface and
also in the non-ohmic/ohmic interface in the limits of perfect
conducting and very low conducting ohmic fluid. We also found that
for intermediate conductivities, the difference between linear
critical values given by the two stability conditions
(\ref{equilibrio}, \ref{Geneq}) is not important (figure
\ref{Tran}). This allows us to assert that the limits of the
stability bands found are actually related to the values
$\sqrt{\varepsilon_l}$ and $\Sigma_0$ and not to the interfacial
total charge change of sign, like stated in the former author's
work \cite*[]{EHD}.

It is interesting to stress that the hydraulic model has the added
value of being simple and intuitive. This has allowed us to find,
for the first time, the stabilizing behaviour of polarization
charges in the ohmic/ohmic interface, even though this type of
interface has been extensively studied (this result is also valid
for the infinite interface). In the non-omhic/ohmic interface, the
model also allowed us to find out the true reason for the
appearance of stability bands in the non-dimensional formulation
in a previous work \cite*[]{EHD}. Thus, the hydraulic model allows
us to find new unknown features and to correct mistaken
interpretations in previously studied systems. The hydraulic model
also sets a reference frame that can be used to find out for what
values of the system parameters (conductivities, ion mobilities,
dielectric constants, relative thicknesses, etc.) the electric
pressure is acting as a destabilising mechanism in a two layer
fluid interface. In addition, the model and its relation to the
standard problem of a long wave perturbation in an infinite
interface, has been set in a very formal and general approach.
This allows for the model to be used as a first step to study a
variety of interfacial (linear and non-linear) stability problems.


\begin{acknowledgments}
I am thankful to Professors Pierre Atten (LEMD-CNRS Grenoble,
France) and A.T. P\'erez (DEE Universidad de Sevilla, Spain) for
fruitful discussion. I also acknowledge financial support from the
Spanish Ministry of Science and Technology (MCyT) under research
project BFM2003-01739 and pre-doctoral research grant FP97
28497673Y.
\end{acknowledgments}

\bibliography{NonLinear}

\vspace{3 mm}

\end{document}